# Dynamic real-time risk analytics of uncontrollable states in complex internet of things systems: cyber risk at the edge

Petar Radanliev[1] · David De Roure[1] · Max Van Kleek[1] · Uchenna Ani[2] · Pete Burnap[3] · Eirini Anthi[3] · Jason R. C. Nurse[4] · Omar Santos[5] · Rafael Mantilla Montalvo[5] · La'Treall Maddox[5]

## Abstract

The Internet of Things (IoT) triggers new types of cyber risks. Therefore, the integration of new IoT devices and services requires a self-assessment of IoT cyber security posture. By security posture this article refers to the cybersecurity strength of an organisation to predict, prevent and respond to cyberthreats. At present, there is a gap in the state of the art, because there are no self-assessment methods for quantifying IoT cyber risk posture. To address this gap, an empirical analysis is performed of 12 cyber risk assessment approaches. The results and the main findings from the analysis is presented as the current and a target risk state for IoT systems, followed by conclusions and recommendations on a transformation roadmap, describing how IoT systems can achieve the target state with a new goal-oriented dependency model. By target state, we refer to the cyber security target that matches the generic security requirements of an organisation. The research paper studies and adapts four alternatives for IoT risk assessment and identifies the goal-oriented dependency modelling as a dominant approach among the risk assessment models studied. The new goal-oriented dependency model in this article enables the assessment of uncontrollable risk states in complex IoT systems and can be used for a quantitative self-assessment of IoT cyber risk posture.

**Keywords** Functional dependency · Network-based linear dependency modelling · Internet of things · Micro-mort model · Goal-oriented approach · Transformation roadmap · Cyber risk regulations · Empirical analysis · Cyber risk self-assessment · Cyber risk target state

## 1 Introduction

This study is focused on standardising the Internet of Things (IoT) risk assessments (Das et al. 2019; Miaoui and Boudriga 2019; Burnap et al. 2017; Radanliev et al. 2020a; Schatz and Bashroush 2017). The contribution of the study is a new goal-oriented dependency model, with the ability to perform dynamic real-time predictive intelligence on threat frequency and the magnitude loss. The aim of the study is to identify a model that enables building dynamic confidence intervals and time bound ranges with real-time data and to address two objectives: First, to identify and capture a target state for cyber risk assessment for the IoT and to adapt a transformation roadmap for existing cyber risk assessments and standards to include IoT risk. For the second objective, the risk quantification is followed by a Goal-Oriented Approach for cyber risk impact assessment through Network-based Linear Dependency Modelling. These are discussed and expanded further in the remainder of this article.

In addition, we also contributed to this topic by introducing a new computational statistical analysis of the literature on the topics of 'Cyber risk and IoT'. We wanted to review and analyse all the literature on this topic, though a qualitative literature review and case study of the related risk assessment approaches and compare the qualitative analysis

✉ Petar Radanliev
  petar.radanliev@oerc.ox.ac.uk

1   Oxford e-Research Centre, Department of Engineering Sciences, University of Oxford, Oxford, UK

2   STEaPP, Faculty of Engineering Science, University College London, London, UK

3   School of Computer Science and Informatics, Cardiff University, Cardiff, UK

4   School of Computing, University of Kent, Canterbury, UK

5   Cisco Research Centre, Research Triangle Park, Durham, NC, USA

with a quantitative analysis of all existing data records on these two topics. By applying quantitative and qualitative approaches, in the text, we wanted to compare different intellectual provenance in the form of study of the intellectual heritage of ideas, concepts, methods, and theories.

### 1.1 Related work

This research follows the guidance of the FAIR Institute (factor analysis of information risk) guidance on quantitative cyber risk assessment (FAIR 2020). This article also aims to address the NIST (NIST 2014) shortcomings in recommendations on how to quantify risk. Although we are not aiming to resolve what NIST has been unable to resolve in decades of research and industry collaboration, we still hope to identify a suitable approach for quantifying emerging cyber risks. This study is guided by the science and practice of resilience (Kott and Linkov 2019; Linkov and Trump 2019) and the existing related work on multicriteria decision in cybersecurity risk assessment and management (Ganin et al. 2017).

## 2 Methodology

Cyber risk traditionally emerges from human–computer interactions (Craggs and Rashid 2017), and the impact analysis result with different estimated loss ranges (Radanliev et al. 2018). The cyber risk investigated in this study emerge from compiling of connected systems, creating risk from data in transit (Anthi et al. 2018) and necessitating standardisation of methods (Tanczer et al. 2018). The methodological approach used for risk quantification in this study is compliant with a Goal-Oriented Approach and Network-based Linear Dependency Modelling. We define goal-oriented approach as the degree to which an organisation focuses on specific tasks and the end results of those tasks. By Network-based Linear Dependency Modelling, we refer to the process of considering relevant network features to explain linear growth, similarly to the recent network-based explanation of COVID-19 linear infection curves (Thurner et al. 2020). Four methodologies have been adapted for IoT risk analysis; those include (a) Risk analysis through functional dependency; (b) risk network-based linear dependency modelling; (c) risk impact assessment with a goal-oriented approach; and (d) integration of the goal-oriented approach with the IoT Micro Mort (IoTMM) model (Radanliev et al. 2018). While there is novelty in combining methodologies that have not been adapted and integrated, the main novelty of this article is the categorisation and assessment of the methodological connections. That differentiates this research from existing quantification models (Radanliev et al. 2020c).

with parameters that are based on expert opinions which can be considered as subjective.

## 3 Bibliographic analysis of literature

We used statistical software to analyse data records on the topic of 'cyber risk and IoT'. To find data records, we did historical analysis (1900–2020) from the Web of Science Core Collection. We identified 191 records of scientific research papers, published in top ranked journals and we analysed the records with R studio. We applied the 'bibliometrix' package in R studio. With the R studio, we designed a three-field plot that enabled us to identify the most prominent keywords, divided by research in different counties—in Fig. 1. To design the three-field plot, we analysed the data from all 191 records combined. The topics of 'cyber risk and IoT' appeared in all data records in combination (jointly). We applied computable statistical analysis to look for further insights on the relationships between 'cyber risk and IoT'. In Fig. 1, we designed a three-field plot, with countries on the left, keywords from the data records in the middle, and areas of focus on the right. We included countries, because we wanted to identify the relationships between the research findings on 'cyber risk and IoT' and compare with the national research efforts. What becomes visible from in Fig. 1 is the higher research output of the US in the keywords that are most present in all data records. Since the USA is the leader in the overall research on 'cyber risk' at present, in Fig. 1 we wanted to determine if the US research is focused on different research areas, and not related to the keywords that are taken as most represented in the combined research records from all countries.

The three-field plot in Fig. 1 identify the USA and UK as the most productive in scientific research on IoT cyber risk. From the three-field plot in Fig. 1, it is not possible to determine if research is conducted by individual, or multiple countries in collaborative projects. To visualise research collaborations, we developed the global collaboration network in Fig. 2. In the Fig. 2 global collaboration network, two different clusters appear (coloured in red + green and blue).

In Fig. 2, we can see how the collaboration lines are developing between the USA, UK, and China, and strong research relationship between Switzerland, Normal, Denmark and Greece. But surprisingly, this analysis shows that UK is not collaborating with the EU partners as strongly as with the USA. In Fig. 2, we created a social structure—collaboration network, with specific countries in the network parameters of the computational program—when designing the Fig. 2 graph. We wanted to evaluate and investigate this result further, and to visualise the type of research topics that separate the two clusters, in Fig. 3 we applied factorial analysis on the same data file, and extracted the keywords and sub-topics in the two clusters. To explain briefly our intentions for applying factorial analysis with computable statistical methods, we wanted to find commonalities by observing the interdependencies between sub-topics. We designed a conceptual structure map with the multiple correspondence analysis (MCA) method. The MCA method is a data analysis technique for nominal categorical data, and we applied the MCA data analysis method to detect and represent underlying structures in the data set. In Fig. 3, the MCA data analysis method represents data as points in a low-dimensional Euclidean space[1].

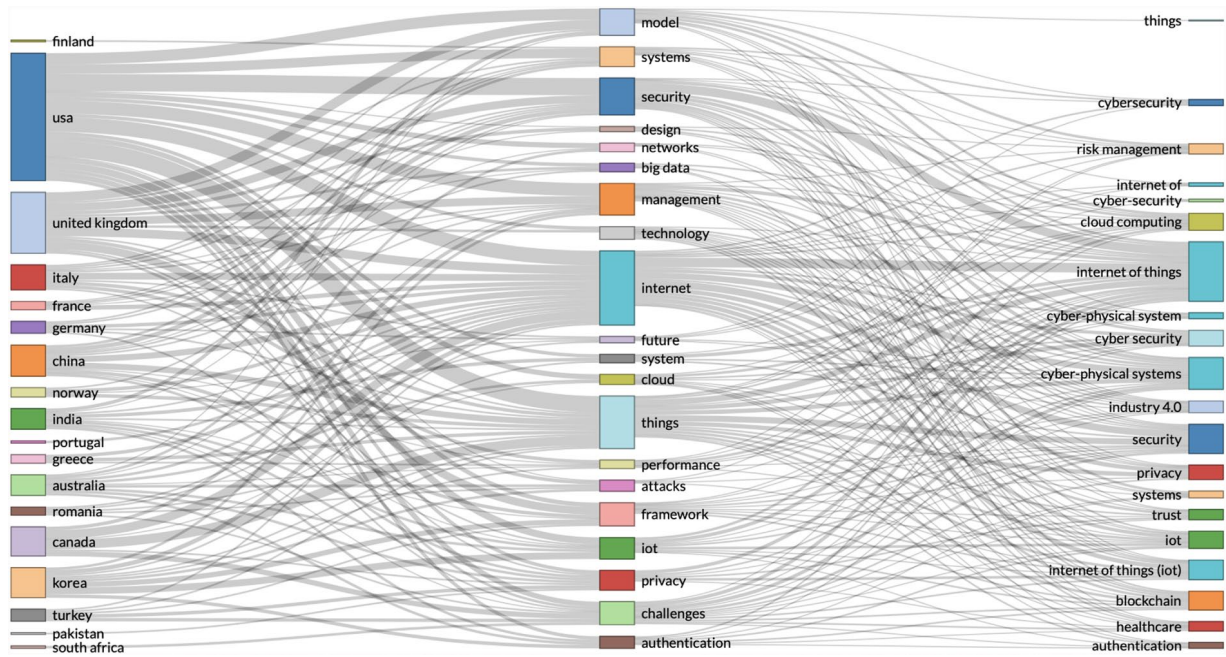

**Fig. 1** Three-Field Plot of sub-topics and keywords in research on IoT cyber risk

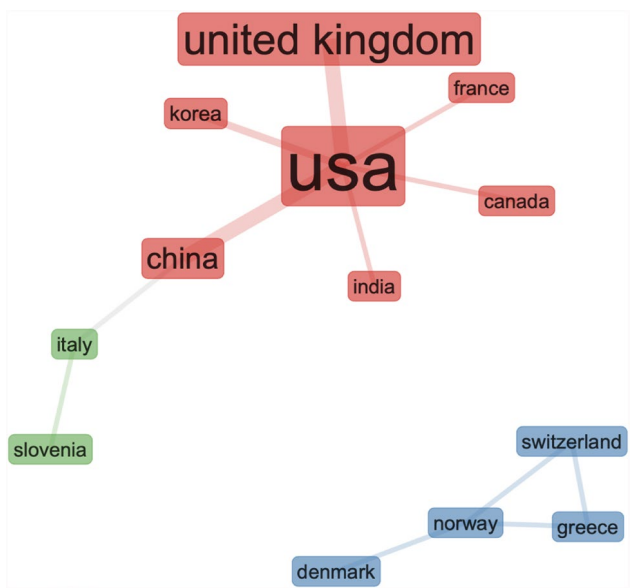

**Fig. 2** Global collaboration network on IoT cyber risk research

We used the factorial analysis statistical methods (in Fig. 3) to visualise the variability in correlated research sub-topics and keywords. The reason we used factor analytic method, was to reduce the keyword variables in the data records and find commonalities by observing the interdependencies between sub-topics. These commonalities guided the case study research on designing a transformational roadmap for moving from current state of risk maturity and reaching a new—target state.

## 4 Case study on transformational roadmaps applied in practice

The bibliometric analysis in the previous section, showed a strong variability in correlated research sub-topics and keywords. By using the factor analytic methods, we reduced the keywords and found commonalities in the interdependencies between sub-topics. With the case study method, we wanted to start building the design process for a transformational roadmap that would enable moving from current state of risk maturity and reaching

---

[1] https://en.wikipedia.org/wiki/Euclidean_space

**Fig. 3** Factorial analysis of research on IoT cyber risk

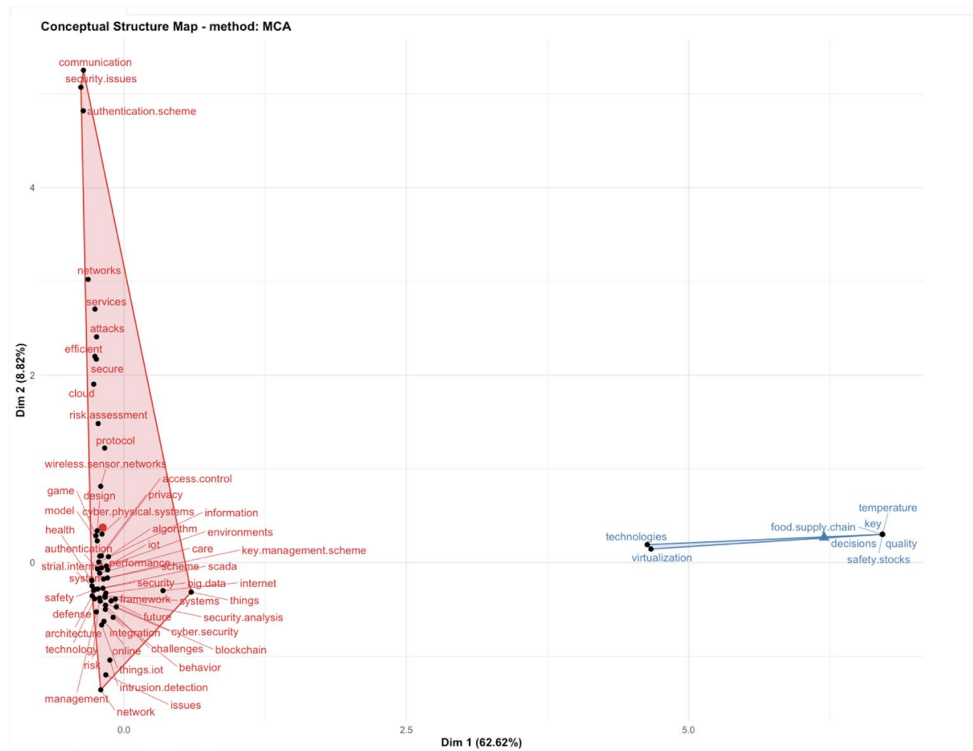

## 4.1 The case study

The case study research involved four workshops that included distinguished engineers from Cisco Systems, and Fujitsu. In the workshops we applied the controlled convergence to verify the design. The controlled convergence was developed in the 1980s by Stuart Pugh (Pugh 1991), using a matrix to compare concepts against a set of pre-determined criteria. The controlled convergence is designed to provide structure to the evaluation of alternative or competing concepts. This approach to pursuing validity through controlled convergence—case study research, follows recommendations from existing literature on this topic (Axon et al. 2018; Eggenschwiler et al. 2016) and provide clear definitions that specify the units of analysis. The reason for pursuing clarity on the units of analysis for IoT cyber risk, was justified by existing literature (Radanliev et al. 2020b), where these are identified as recommended areas for further research (de Reuver et al. 2017). The IoT risk units of analysis are verified with the controlled convergence method, where experts were asked to confirm the valid concept, merge duplicated concepts, and delete conflicting concepts.

a new—target state. In the case study, we have selected the most prominent cyber risk assessment methods, and we attempted to find commonalities interdependencies between different risk assessment approaches.

## 4.2 Transformation implementation tiers for reaching the target state based on a current state—categorised with a Goal-Oriented approach

Creating a connection between different independent cyber risk models and IoT risk is difficult, because cyber risk assessments are not based on standardised risk estimation. To standardise the IoT risk estimation, we focused on the success factors and dependencies with a goal-oriented modelling and Bayesian methods.

The previous steps are applied for the identification of a target state and the controlled convergence is applied for the development of a transformation roadmap, specific for the case study scenario. To build the transformation roadmap, firstly we identified the specific target state implementation tiers. Secondly, the specific implementation tiers and their relation to the main cyber risk impact assessment approaches is validated with the controlled convergence. The rationale is that different scenarios will have different implementation tiers to perform in order to transition to a higher maturity lever. A long and detailed list of implementation tiers can be found in the NIST framework and the Exostar system.

In Table 1 we describe with examples the process of building a transformational roadmap. The process is based on determining implementation tiers in the form of 'parent and child' or 'goal and objective'. Building the transformational roadmap with case specific implementation tiers,

**Table 1** Transformation implementation tiers categorised with a Goal-Oriented approach—describing how the transformational roadmap can be applied in a case-specific scenario

**Transformation roadmap with case specific implementation tiers**

**Training and awareness**

Control goal (**parent**) 1: Security skills assessment and training for IoT systems

Control objective (**child**) 1: Skills and integrated plan to support defence of IoT systems

Control element (**orphan**) 1: Analysis of needed skills; provide training to match the required skills and validate skills through periodic tests. More advanced control orphans include: security assessments using real-world examples to measure mastery or skills

Control goal (**parent**) 2: Penetration testing of IoT systems

Control objective (**child**) 2: Test the defences by simulating IoT cyber-attacks

Control element (**orphan**) 2: Regular focussed penetration tests for detecting unprotected systems through vulnerability scanning and penetration testing combined

Control goal (**parent**) 3: IoT risk from mobile device

Control objective (**child**) 3: Mitigate cyber risk from mobile devices

Control element (**orphan**) 3: Mobile devices should have access controls to enforce policies and option to remotely clean the device

**Cyber threat intelligence**

Control goal (**parent**) 1: IoT boundary defence

Control objective (**child**) 1: Manage the flow of information between network trust levels

Control element (**orphan**) 1: Prevent communications with malicious IP addresses, use two-factor identification; design DMZ network and scan connections that aim to bypass the DMZ; block known bad signature or attack behaviour

**Security event monitoring**

Notes: links with: (a) network security; (b) identity and access management

Control goal (**parent**) 1: Maintenance, monitoring and analysis of IoT audit logs

Control objective (**child**) 1: Collect, manage and analyse IoT audit logs of events

Control element (**orphan**) 1: Two synchronised timestamps in logs to ensure consistency; develop IoT log retention policy

Control goal (**parent**) 2: Secure configurations for IoT networked devices—such as firewalls, routers and switches

Control objective (**child**) 2: Actively manage the security configuration of the IoT network infrastructure

Control element (**orphan**) 2: Documenting all new IoT configurations rules that allow traffic to flow through network security devices; use two-factor identification and encryption

Control goal (**parent**) 3: Account monitoring and control of IoT data

Control objective (**child**) 3: Control the life cycle of IoT system and IoT devices applications accounts

Control element (**orphan**) 3: Disable unused IoT devices accounts; imprint accounts expiration date; enable revoking system access for IoT devices accounts; log-off IoT devices after a standard period of inactivity; encrypt transmitted passwords for IoT devices enables the application of the goal-oriented approach. The roadmap in Table 1 describes this process. Worth emphasising that in Table 1 we present the transformational roadmap as a 'parent, child and orphan' methodology that is required for the dependency modelling. Using the case study research, to build a 'parent, child and orphan' table as a summary map of implementation tasks (tiers), by applying the controlled convergence evaluation matrix, enables the categorisation of concepts as a simple 'understanding' of a task (tier), without any data to back up that understanding, or as a established concept, based on evidentialist understanding. These are categorised in groups, and numbers, then colour coded for visibility and ease of understanding—in Table 1.

The process in Fig. 4 enables transforming a 'parent' statement, which represents 'understanding' and 'information', into an 'orphan' statement describing a measurable outcome and therefore enables collecting probabilistic data for the 'justification of truth'. The statements are then compared to the risk assessment approaches (Caralli et al. 2007;

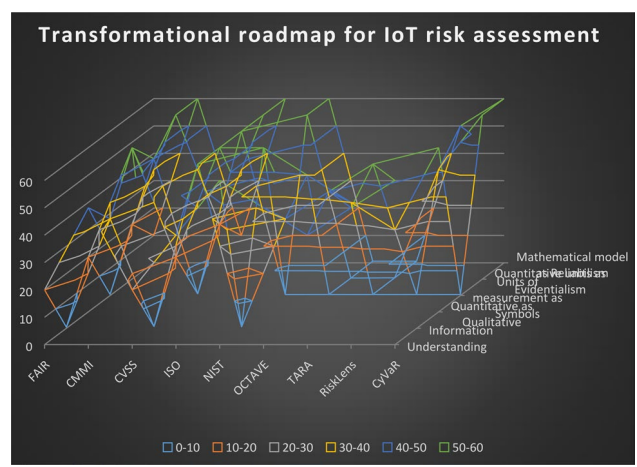

**Fig. 4** Transformation implementation tiers categorised in Microsoft Excel to reflex the level of the 'justification of truth'—simple understanding vs evidentialism

CMMI 2017; CVSS 2019; Cyberpoint LLC, n.d.; FAIR 2020; ISO 2017; NIST 2014; Wynn et al. 2011), which are further evaluated in Sect. 5.

When applying the implementation tiers of the transformational roadmap—described with Microsoft Excel in Fig. 4, the 'parent, child and orphan' implementation tiers will be almost inevitably different, because the tiers are always case specific. The statements in the categories represent examples for clarifying the methodology. For a more detailed process on implementing the recommendations emerging from the transformational implementation tiers, we refer to the NIST cyber security implementation tiers as support guidance. The implementation tiers in Fig. 4 advance the NIST implementation tiers by advocating a methodology for transforming current cyber profiles that dictate reactive approach, into risk informed, repeatable and adaptive target cyber profiles (Barrett et al. 2017). The difference in our methodology is that the target state implementation tiers are focused on identifying case specific IoT risk, rather than examining general cyber risk categories and subcategories. Hence, we addressed a significant gap in the NIST cyber risk assessment process. The following section details how the implementation tiers are applied in for IoT risk impact assessment with a dependency goal-oriented modelling.

# 5 IoT risk assessment with dependency goal-oriented modelling

We applied dependency goal-oriented modelling to connect the implementation tiers in the IoT risk assessment. The first step is to link separate implementation tiers. This requires identifying the shared principles from the tiers being connected. Then, to determine the level of dependency risk, it is necessary to understand the dependencies of the shared principles that are representative of a larger complex IoT system.

## 5.1 IoT risk impact assessment with a Goal-Oriented Approach

Our methodology for IoT risk impact assessment with a goal-oriented approach focuses on the success factors and concentrates on the external dependencies. In this approach, individual IoT risks are considered as representative of a larger complex IoT system. This advocates a top-down or goal-oriented modelling approach, where success factors are traversing across multiple isolated models. The proposed goal-oriented modelling approach is analysed with Bayesian methods. Statistical inference can be updated with Bayes theorem as more evidence and probabilistic data become available. The paradigm can provide a real-time statistical/probabilistic assessment of IoT cyber risk of all the entities in the model. The dependencies between a dynastic metaphor, such as 'parent', 'child', and 'orphan' as explained in the transformational implementation tiers (Table 1), can be analysed with computational statistics using a Bayesian analysis engine (Hanson and Cunningham 1996; Weinberg, n.d.). Therefore, the proposed goal-oriented dependency modelling approach relies on Bayesian analysis engine to evaluate a range of sensitivities or vulnerabilities. To develop Bayesian inferences, the model requires the statistical relationships between parent, child and orphan nodes (where available) (see Table 1). The level of achievement is articulated as the 'state' of the goal, which is usually correlated with failure and success. However, while 'state' is usually identified according to a value judgement, e.g. bad-good, no-yes or 0–1, a 'state' could also be identified with a number of possible states, not necessarily two single states. In addition, with the IoT real-time update feature, the decision model can provide real-time risk assessment, where the impact of changes in state can be immediately identified through the functional dependencies.

## 5.2 IoT risk analysis through functional dependency

Dependency modelling and analysis provides a means to support the management of functional and operational complexities within IoT systems with focus on the system elements, measures of a design or operational challenge, and the functional dependencies that define their associations. Dependency modelling and analysis can support a superior understanding of connectivity and its implications on performance, and can assist in constructing, improving, and maintaining such complex systems.

The construct and exchanges that happen in IoT domain defines a tightly coupled association amongst constituting components and sub-systems that typically rely on the correct functions of another linked component or node. This exhibits a dependency relationship, which can either be direct (a first order dependency) or indirect (a subsequent higher order dependency) (Laugé et al. 2015) from both layer and component-level perspectives.

Dependency modelling and analysis provides a means to support managing functional and operational complexities in IoT systems with focus on the system components, measures of a design and/or operational challenge, and the functional dependencies that define component associations. In an IoT ecosystem, component/system functionalities are typically reliant on connectivity or network exchange/communication infrastructures—mainly the internet, and service functionalities and availabilities of connected IoT components (Yadav et al. 2019). From an availability standpoint, components functionalities and the service(s) they enable can be greatly hindered by intentional or unintentional disruptions in network connectivity. This is especially when the disruption involves a top-level information processing and distribution node like a wired/wireless router. Affected devices at the lower end of the network can thus be impacted to the degree with which they rely on service receptions from a connected service/function-impaired component.

For example, from a typical IoT layered architecture (Bilal 2017), components, functions or services on the *'application layer'* typically rely on the normal functioning of linked components/nodes/services on the *'network layer'*. In turn, the network layer components/services rely on inputs from the *'perception layer'*. Compromising, disrupting or destroying components/service operations on a higher-level layer, e.g. perception, can result in wrong or no information exchange via connectivity. Thus, the accurate functioning of connected components/service(s) on the network or application layers can be altered. This suggests that the broader security risks in IoT domain (like any other independency system) may not be entirely drawn from the failure of one specific IoT component. Most often, it extends to the failure of other linked components/services that can receive cascading impacts. This dependency amongst IoT sub-systems and components can worsen when adverse impacts flow from one affected component/system/service onto another (Bloomfield et al. 2010; Kotzanikolaou et al. 2013).

Dependency modelling and analysis can support a superior understanding of component connectivity and implications on performance. It can also assist in constructing, improving, and maintaining of such complex system. Typically, security and safety–critical impacts can vary amongst assets, their functionalities (services), placement positions, and configurations within industrial networked systems (Ani et al. 2018) including IoT. However, effective control and management decision-making can be supported by adopting risk assessment methods that go beyond considering risk scenarios one by one, qualitatively or statically, to considering the relationship between the risk factors. This way, security-related dependencies can be evaluated such that can enable profounder insights on how an impact to IoT infrastructure that prevents it from delivering the relevant and requisite service(s), can affect the performance levels of other sub-systems connected and reliant on an affected target. This can support adopting effective security incidence response and recovery (Laugé et al. 2015), and help minimise the effects of IoT disruptions.

From a goal-oriented perspective, an effective approach involves evaluating the security risks of an IoT component/device or service being compromised or failing due to the compromise and(or) failure of another component or service(s) the first entity relies upon. The goal of ensuring or maintaining operational continuity makes crucial to determine and understand the extent or number(s) of dependencies necessary to sustain continuity. Thus, the security risks that can cause security and functional impairments in IoT components/services can be characterised by a chain of cascading failure-causing impacts. These when understood earlier can support relevant security strategy adoption as solutions. Earlier understanding of impact flows and operational implications can also help in identifying potential critical points in the network, and support prioritisation of risk management to ensure that more critical security risks are addressed first.

## 5.3 Network-based linear dependency modelling

There are a number of approaches for modelling complex system dependencies which provide a good starting point for exploring dependency analysis in IoT. The Leontief-based input–output approach (Setola et al. 2009) can support IoT interdependency modelling from a failure perspective. The input–output Leontief-based model offers an explorable way to derive dependency-oriented attributes such as; disruption probability, risk transmission, and cascading impacts (Zhang and Peeta 2011). A scenario-driven what-if analysis method enables a simulation-based technique for evaluating the consequences of discrete events and physical economic flows amongst IoT sub-systems (Zhang and Peeta 2011), enabling the resolution of varied events and controls strategies. However, the more common approach involves the network-based approach, which can assist with evaluating physical dependencies and cascading disruptions within and across geographical dimensions. These typically explore stochastic modelling (Bloomfield et al. 2010; Huang et al. 2016), especially the Markov-based undirected techniques (Nozick et al. 2004; Qiao et al. 2007), and the Bayesian Network-based (BN) technique (Di Giorgio and Liberati 2011). This BN approach appears more promising for demonstrating the characteristics of components and dependencies in the IoT network, thus discussed further. It is crucial to note the common denominator that the listed approaches leverage topological attributes of networked infrastructures, such as connectivity, path length, degree of vertex, and redundancy ratio for interdependency analysis and the resolve of spatial impacts of security risks and disruptions in the IoT.

Bayesian Networks quite well supports IoT dependency modelling through functionality evaluations, which involves analysing how the functionality of one system or component can affect the functionalities of other systems or components (Zhang and Peeta 2011) on the basis of connectivity and process configurations over multi-layered IoT architecture. Building on graph theory (Laugé et al. 2015; Stergiopoulos et al. 2016), BN's (Di Giorgio and Liberati 2011) graphical representation of probability distributions provide a useful dependency analysis features akin to the IoT, i.e. consisting of a set of discrete or continuous random variables: $V = \{v_1, v_2, v_3, \ldots, v_n\}$. BN enables a directed acyclic graph where vertices/nodes are marked with quantitative probability information, such that:

1. For each random variable, there exists an associated vertex (representing network components/nodes) in the BN.
2. Edges (representing links) use arrows to represent the relations among random variables, so that an arrow $v_i, (i = 1 - n, corresponding to the number of parent nodes)$ from a node $v_i$ to another node $v_{i+}$ in the set V implies that the state of $v_i$ directly influences the state of $v_{i+}$, which in turn influences a node $v_{i++}$ (herein $v_i$ is considered a parent, $v_{i+}$ is a child node).
3. For each node (corresponding to a component or service), parent's influence can be evaluated using a conditional probability distribution (CPD): $P(v_i/Parents(v_i))$

The CPD for parents' influence would often work well for discrete random variables/nodes; however, the potentials for random nodes to increase introduces intractable complexities (Di Giorgio and Liberati 2011) that require that certain conditional independence relations be encoded in the BN graph to simplify complexities (Jensen 2007). Basic BN's inability to consider time attributes makes it weak in accounting for potential changes system while evaluating influences and dependencies. Dynamic Bayesian Network (DBN) resolves this by enabling a modelling of systems that evolve in discrete time steps called 'Time slice Bayesian Network'—TSBN. TSBN links to 'time slices' and provides a static model of the system at each instant of time, which instinctively reflects rapid causalities. Dynamics are then captured by linking inter-time slices, also indicating temporal probabilistic dependencies between random nodes belonging to varied time slices.

Supposing that the condition guiding the system development remains unchanged, and DBN can be defined by topologies of sequential time slices, then it becomes feasible to; infer the system's current status learning from available past and present information, estimate the past status and predict future status of the system. Reiterating the point from (Di Giorgio and Liberati 2011), it becomes feasible to evaluate the probability distribution using information aggregated into a conditional probability table, if nodes in the network remain ordered such that a node's index is always preceded by the indices of its parent. The probability distribution for the network of nodes V is thus;

likening the typical IoT layered architecture (comprising perception, network, and application layers) to traditional service-oriented structures, where the occurrence of service disruptions/failures is represented at the lowest level of the TSBN. Similar to the IoT, TSBN service structure and positioning is divided into three different levels: Atomic Events, Propagation, and Service (Di Giorgio and Liberati 2011). This structure points that, typically events in a TSBN dependency interrelationship expresses these three structures at some point and degree.

Atomic events level refers to the part of a TSBN event that captures random nodes/components associated with adverse events capable of causing compromise, disruptions and impairments (Di Giorgio and Liberati 2011). Essentially, it indicates the potential starting point of an abnormal activity in the network, which can occur at any layer of the IoT system depending on the target or victim component/node of compromise. Appropriate security respond and management need to account for potential risk attributes (threats, vulnerabilities, and attack likelihoods) from identified or associated event origins as measure towards grasping the full scope of impacts and criticalities of events. As component/node functional dependencies move in a particular direction from event origin, typically, the flow of probability impacts for nefarious events would move in the opposite direction.

The propagation level is associated to nodes through which service functions are enabled based on connectivity (Di Giorgio and Liberati 2011). This also indicates the medium through which potential negative impacts are enabled and trickle through the network of connected IoT. Thus, it includes facilitating interconnections of nodes and services to support the attainment of desired system goals, and can emerge at any point in the IoT layered structure, provided that affected components/nodes support the exchange or transfer of services to certain ends. Evaluating associated probabilities of security risks at this level can enable insights into the scope of feasible negative impacts that need to be prepared against or responded to.

The service level is associated to nodes through which the final desired function of an IoT is achieved. It can also refer to the point of goal accomplishment for the system or user (Di Giorgio and Liberati 2011). It can also be associated to the probable final point of negative impacts from an initial atomic event through a propagation level based on

$$P(v_1, v_2, v_3, \dots, v_n) = \prod_{i=1}^{n} P(v_i | v_{i-1}, v_{i-2}, \dots, v_1) = \prod_{i=1}^{n} P(v_i | Parents(v_i))$$

The above DBN theory may be applied to IoT system/infrastructures interdependency analysis to support security risk management. The goal-oriented approach enables first or multi-order links and dependencies. Understanding the probable security risk to and at this level helps to achieve a more holistic viewpoint of the full scope of the impact of a security event in the network, which is necessary to inform an effective solution approach.

**Fig. 5** Network-based (Component/Layer) Linear Interdependency Structure

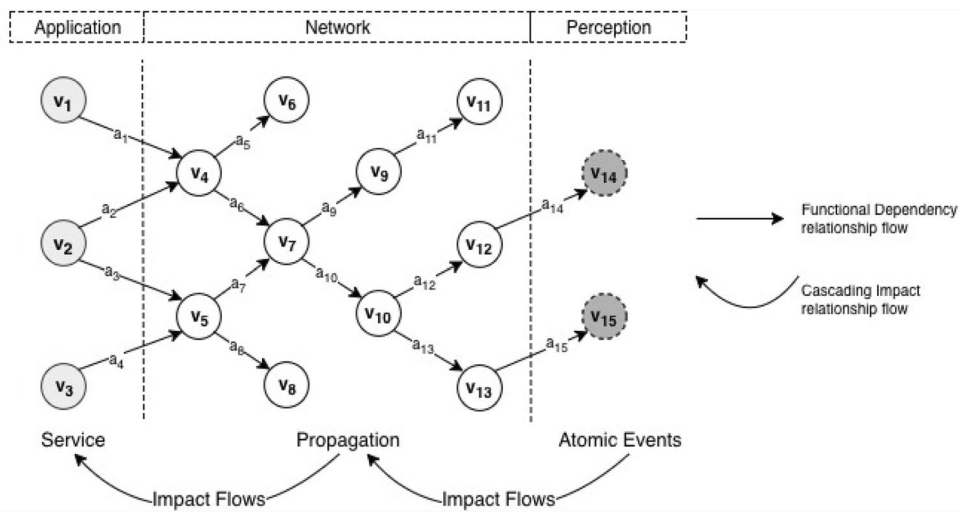

Using the IoT layer architecture as example, the standard functions for components and services on the application layer (which may be considered the final goal level of component functions) in an IoT system, typically depend on the standard functioning of movement of data through the network layer, and unto the perception layer (PL) (which perceives the physical properties of things). Using Fig. 5 if a service flow $a_{14}$ on a perception layer is impaired, it can affect the functionality of a connected services $a_{12}, a_{10}, a_7, and a_6$ on the network, and $a_4, a_3, a_2, and a_1$ on application layer. The services on the application layers result in the final goal, and their negative impacts on disruptions or failure in attaining the final goal.

Combining all three service contexts in relation to abnormal events in IoT networks can provide a useful way if representing functional dependencies to support security risk management using the goal-oriented approach. In addition, functional dependency assessments may be combined with other dynamic vulnerability information such as the environmental severity attributes (*vulnerability collateral damage potential*, its *target distribution*, and its *target security requirement violation*) often viewed as optional from standard CVSS models (Chejara et al. 2013; Mell et al. 2007) when undertaking security analysis. The outcome can potentially be used to understand potential impacts associated to component vulnerabilities—*estimating how wide or otherwise an impact on a component or system can ripple through and affect other components and systems*. Such insights can support a more accurate estimation of security risks and cascading impacts, and plan incident response and recovery in the IoT domain. Thus, an organisation embracing IoT into its network need to perform such multi-factor security analysis to better grasp its IoT cyber security posture. Organisations adopting IoT like any other digital trend need to be clear about their current and target security posture, before designing and adopting a transformation roadmap outlining tasks to achieve the stated target posture. In developing a target profile, a broad range of approach may be used, considering more effective and efficient risk management approaches across the entire in-scope organisations.

### 5.4 Discussion on the advantages of applying the Bayesian network method

The original idea of this article was to advance the efforts of other quantitative risk assessment approaches, such as the factor analysis of information risk (FAIR 2020). Similarly, the System Theoretic Process Analysis (STPA) method is also used for identifying the risks and vulnerabilities in the cyber-physical system from the safety and security point of view. The main novelty of this article is that the proposed method does not only consider the risks caused by individual components, but also considers risks caused due to misinteractions between different components. A similar attempt was made in a recent study to address this issue with applying Monte Carlo (MC) simulation with the FAIR method, which was supported by the RiskLense software. It was found that the Bayesian network method achieves higher accuracy in cases that cannot be accurately modelled by the FAIR model (Wang et al. 2020). Moreover, the Bayesian network method is more flexible and extensible by showing how it can incorporate process-oriented and game theoretic methods. Presenting the potential for an integrated cybersecurity risk assessment.

### 5.5 Discussion on limitations and further research

This study compares a number of most prominent risk assessment models, methods and frameworks. However, a holistic analysis of all risk assessment approaches was considered beyond the scope of this study. Hence, it is difficult to claim that the approach we present, will be compatible with existing, and new approaches. However, the flexibility of our approach is that we applied epistemological approach, to the Goal-Oriented Approach and Network-based Linear Dependency Modelling. The epistemological approach in this article, presents a generic approach that can easily be adapted to accommodate additional concepts and to guide researchers and practitioners on the advancement of this topic.

### 5.6 Discussion on further research—the relationship between the Goal-Oriented Approach and the IoTMM model

The relationship between the IoTMM and the Goal-Oriented Approach is that the IoTMM expresses a conditional probability for the Goal-Oriented Approach dynastic metaphor. The conditional probability is the probability of a goal in the dynastic metaphor (in the transformational roadmap), being in a 'state' according to the states of its 'parents', 'children', and 'orphans'. Since Bayesian inferences are based on a logical AND or OR function then the IoTMM provides more clarity than a simple structure consisting merely of 0 s and 1 s.

In scenarios where there is a lack the probabilistic data to determine the IoTMM, then the goal is considered as 'uncontrollable'. In such scenarios, the IoTMM still provides a conditional probability as an isolated model. However, in uncontrollable scenarios, the IoTMM only presents a catalogue of the probabilities in each possible 'state'. The advantages of combining the IoTMM with dependency modelling is more obvious when the uncontrollable states are dynamically analysed from other distributed states that contain the actual 'state' and dependencies. By using the two approach simultaneously, it becomes possible to assess uncontrollable states in complex systems.

This ability to assess uncontrollable states in complex systems can be used as a decision-making method. In this paper we established that cyber risk regulations for the IoT do not exist. This creates invisible risks and also triggers data protection questions from the new types of cyber risk. In this paper we also conclude that the IoT risk is not included in the cyber security assessment standards, hence, often not visible to cyber security experts. This is concerning, because companies integrating IoT devices and services need to perform a self-assessment of its IoT cyber security posture. Although there are emerging cyber-security standards related to IoT (e.g. IEC62443 (Shaaban et al. 2018), the NIST Guide to Industrial Control Systems

(ICS) Security[2], or the ENISA Good Practices for Security of Internet of Things[3]), this article presents a new goal-oriented approach, that provides a transformation approach for reaching the required or desired target state of cyber security/risk maturity level.

Through epistemological analysis we uncovered the best method to design an IoT cyber risk assessment. We also presented and evaluated a transformation roadmap for IoT cyber risk assessment with a case study research, the controlled convergence, and the goal-oriented approach. The process we presented enables practitioners to improve their cyber posture. The new understanding of how to assess IoT risk dependencies also enables the development of new cyber regulations.

## 6 Conclusion

This article reviews the existing literature and performs comparative, empirical and epistemological analysis of common cyber risk assessment approaches and integrates current standards. The findings present a map of the present initiatives, frameworks, methods and models for assessing the impact of cyber risk. Hence, the article advances the efforts of integrating cyber risk standards and governance and offers a better understanding of a holistic assessment approach for IoT cyber risk. This enables visualising the interactions among different sets of cyber security assessment criteria and results with a new design criterion specific for cyber risk from the IoT. The contributions from this paper constitute the following:

1. Transformation roadmap for IoT cyber risk assessment and
2. Dependency modelling describing how IoT companies can achieve their target state with goal-oriented transformation implementation tiers, that can be applied for the following:
a) Risk identification (measure IoT risk);
b) Risk management (standardise IoT risk);
c) Risk estimation (compute IoT risk) and
d) Risk prioritisation (design IoT risk strategy).

New methods are presented for the following:

1. IoT risk analysis through functional dependency, providing a superior understanding of connectivity and its

implications on performance, assisting in the construction, and maintenance of complex IoT systems;

2. Network-based linear dependency modelling, which enables evaluating physical dependencies and cascading disruptions within and across geographical dimensions. Hence, this approach appears more promising for demonstrating the characteristics of components and dependencies in the IoT network;

3. IoT risk assessment with a goal-oriented approach, and with the IoT real-time update feature, the decision model can provide real-time risk assessment, where the impact of changes in state can be immediately identified through the functional dependencies and

4. A correlation between the goal-oriented approach and the IoTMM model for assessing uncontrollable states in complex systems. The uncontrollable states are dynamically analysed from other distributed states. By using the two approaches simultaneously, it becomes possible to assess uncontrollable states in complex systems, which can be used as a decision-making method.

These findings are relevant to national and international digital strategies, specifically for IoT cyber risk planning.

**Acknowledgements** Eternal gratitude to the Fulbright Visiting Scholar Project.

**Author contributions** Dr PR: main author; Prof. DDR and Prof. MVK: supervision; Dr RMM, OS, LM, Prof. PB, Dr JRCN and Dr UA, EA: review and corrections.

**Funding** This work was funded by the UK EPSRC [Grant No: EP/S035362/1] and by the Cisco Research Centre [Grant No. 1525381].

**Data availability** All data and materials are included in the article.

## Compliance with ethical standards

**Conflict of interests** On behalf of all authors, the corresponding author states that there is no conflict nor competing interest.